\documentclass[12pt]{article}
\usepackage{graphicx}
\usepackage{subfig}

\newcommand\pubnumber{SNSN-323-63}
\newcommand\pubdate{\today}

\def\institute{}

\def\Title#1{\begin{center} {\Large #1 } \end{center}}
\def\Author#1{\begin{center}{ \sc #1} \end{center}}
\def\Address#1{\begin{center}{ \it #1} \end{center}}

\newcommand\pubblock{\rightline{\begin{tabular}{l} \pubnumber\\
         \pubdate  \end{tabular}}}
\newenvironment{Abstract}{\begin{quotation}  }{\end{quotation}}
\newenvironment{Presented}{\begin{quotation} \begin{center} 
             PRESENTED AT\end{center}\bigskip 
      \begin{center}\begin{large}}{\end{large}\end{center} \end{quotation}}

\def\beq{\begin{equation}}
\def\eeq#1{\label{#1}\end{equation}}
\def\eeqn{\end{equation}}

\def\beqa{\begin{eqnarray}}
\def\eeqa#1{\label{#1}\end{eqnarray}}
\def\eeqan{\end{eqnarray}}

\let\bar=\overbar

\def\Dslash{\not{\hbox{\kern-4pt $D$}}}
\def\dslash{\not{\hbox{\kern-2pt $\del$}}}

\def\msb{{\bar{\ssstyle M \kern -1pt S}}}

\begin{document}
\begin{titlepage}
\pubblock

\vfill
\Title{Charge asymmetry measurements in $t\bar{t}$ events at the LHC}
\vfill
\Author{ Mohsen Naseri 
\\
on behalf of the ATLAS and CMS Collaborations}
\Address{\institute School of Particles and Accelerators, Institute for Research in Fundamental\\
Sciences(IPM),\\
P. O. Box 19 56 83 66 81, Tehran, Iran
}
\vfill
\begin{Abstract}
An overview of the most recent measurements on top quark charge asymmetry in top quark pair production is presented. The results are obtained using data collected with ATLAS and CMS detectors in proton-proton collisions at centre-of-mass energies of 8 TeV. In these studies, either dileptonic or semileptonic top pair decays are analyzed. All measurements are comparable with the standard model prediction and no sign of new physics is observed.
\end{Abstract}

\vfill
\begin{Presented}
$9^{th}$ International Workshop on Top Quark Physics\\
Olomouc, Czech Republic,  September 19--23, 2016
\end{Presented}
\vfill
\end{titlepage}
\def\thefootnote{\fnsymbol{footnote}}
\setcounter{footnote}{0}

\section{Introduction}
Top quarks are produced in pairs via the strong interaction either from the fusion of two incoming gluons or from the annihilation of an incoming quark and anti-quark at leading order (LO). The interference effects between the Born diagram and the box diagram at next-to-leading order (NLO), as well as between initial- and final-state radiation, correlates the direction of the produced top quarks(antiquarks) to the direction of the incoming quarks(antiquarks).  

At the LHC,  quarks in the initial state are mostly valence quarks while the initial state antiquarks are always from the sea quarks. As valence quarks on average carry a harder momentum spectrum than sea-quarks, top quarks have a preference to be produced in forward directions. The rapidity distribution of top quarks is therefore predicted to be broader than that of the more centrally produced top antiquarks. The difference of the absolute values of the rapidity of top quark and antiquark in an event, $\Delta|y|$, can be used to measure the asymmetry effect.
\section{The $t\bar{t}$ charge asymmetry in lepton+jets events}
\begin{figure}[htb]
\begin{center}
     \subfloat[]{\includegraphics[width=0.48\textwidth , height=0.35\textwidth]{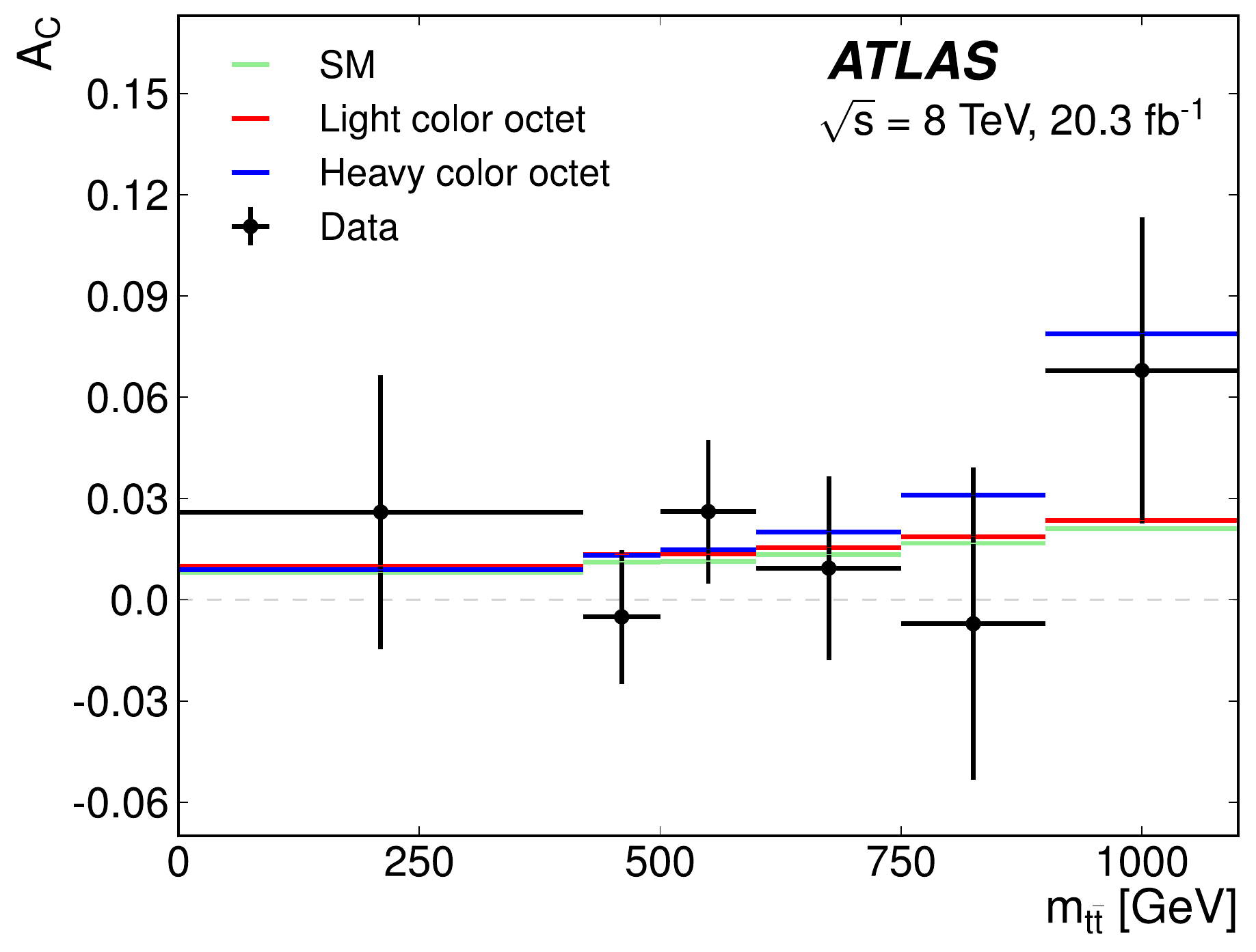}}
     \subfloat[]{\includegraphics[width=0.5\textwidth , height=0.38\textwidth]{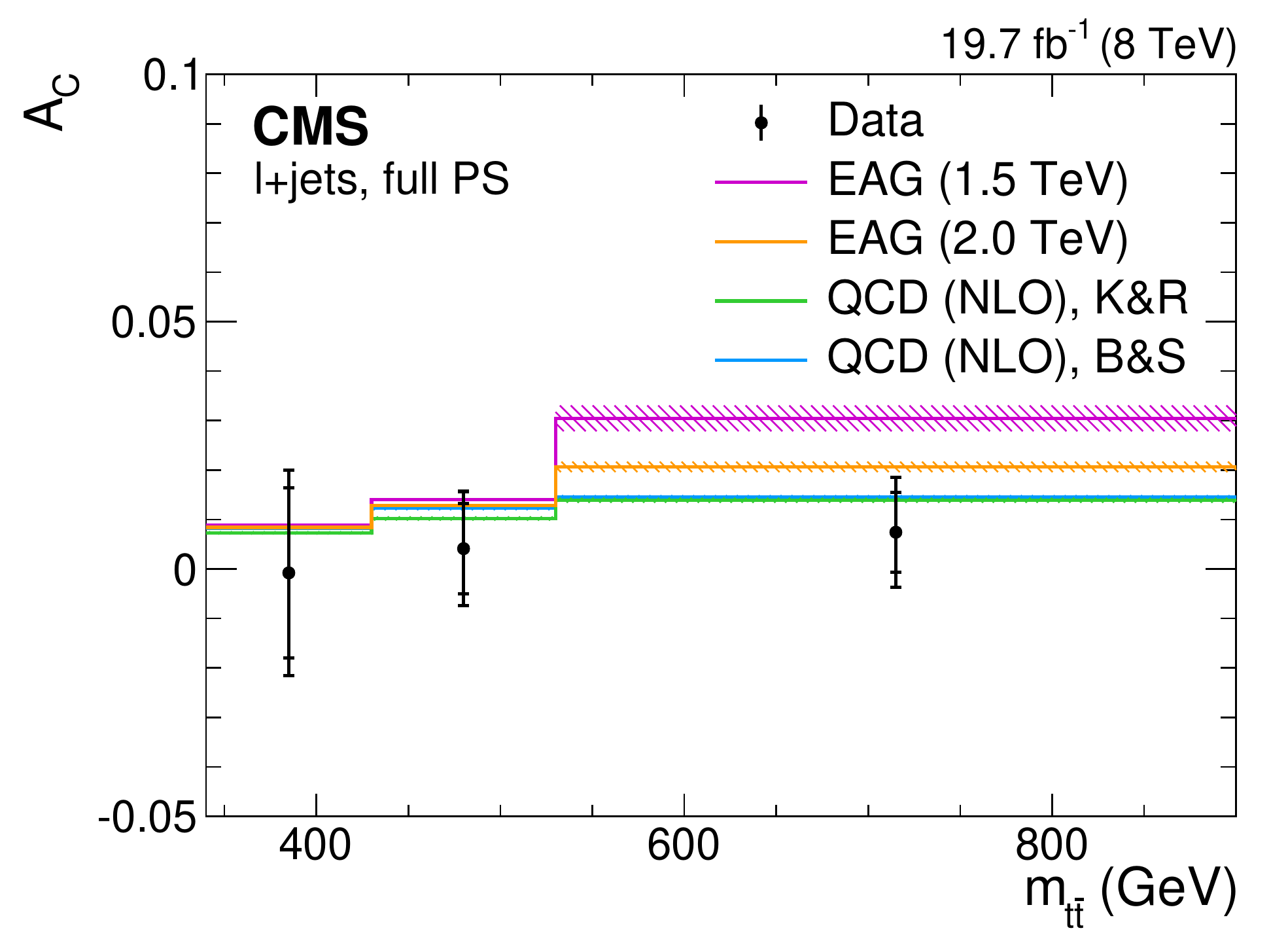}}
     \caption{(a) Measured asymmetry as a function of $m_{t\bar{t}}$ in ATLAS~\cite{Aad:2015noh}, and (b) in CMS ~\cite{Khachatryan:2015oga} using lepton+jets events, compared to predictions for the SM, as well as to the predictions of BSM in different sectors.\label{fig::semi}}
\end{center}
\end{figure}

The CMS collaboration\cite{Chatrchyan:2008aa} performs the inclusive and differential measurements of the charge asymmetry in lepton+jets events using the data corresponds to an integrated luminosity of 19.7 fb$^{-1}$~\cite{Khachatryan:2015oga}. The observed distributions of  $\Delta|y|$ are corrected back to parton level by imposing an unfolding technique to allow for a comparison of the measurement and the prediction from theory. 
The differential charge asymmetry is also measured as function of kinematic variables of the $t\bar{t}$ system, the invariant mass $m_{t\bar{t}}$, the absolute value of the rapidity $|y_{t\bar{t}} |$, and the transverse momentum $p_{T,{t\bar{t}}}$ of the $t\bar{t}$ system. Figure~\ref{fig::semi} shows the differential results as a function of  $m_{t\bar{t}}$. The inclusive measurement yields an asymmetry of A$_{C}$ = 0.0010 $\pm$ 0.0068(stat.)$ \pm$ 0.0037(syst.).  As an alternative, the measurements are also performed in a fiducial phase space, yielding an integrated result of  -0.0035 $ \pm$ 0.0072 (stat.) $\pm$ 0.0031 (syst.).

The CMS collaboration also reports the most precise measurement of the inclusive charge asymmetry using a template fit to $\Upsilon_{t\bar{t}} = \tanh\Delta|y|$, as sensitive variable~\cite{Khachatryan:2015mna}. The measured charge asymmetry is A$_C$ = 0.0033 $\pm$ 0.0026(stat.) $\pm$ 0.0033(syst.), which is consistent with SM predictions.

The ATLAS collaboration\cite{Aad:2008zzm}  analyzes data corresponding to an integrated luminosity of 20.3 fb$^{-1}$~\cite{Aad:2015noh}. To measure the charge asymmetry, the observed  $\Delta|y|$ distribution is unfolded to parton level using the Fully Bayesian Unfolding(FBU) technique. In addition to the inclusive measurement, the differential measurements are performed as a function of the invariant mass, transverse momentum and longitudinal boost of the  $t\bar{t}$ system. The inclusive $t\bar{t}$ charge asymmetry is found to be A$_C $ = 0.009 $\pm$ 0.005.
\section{The $t\bar{t}$ charge asymmetry in dilepton events}
\begin{figure}[ht]
\begin{center}
     \subfloat[]{\includegraphics[width=0.4\textwidth , height=0.3\textwidth]{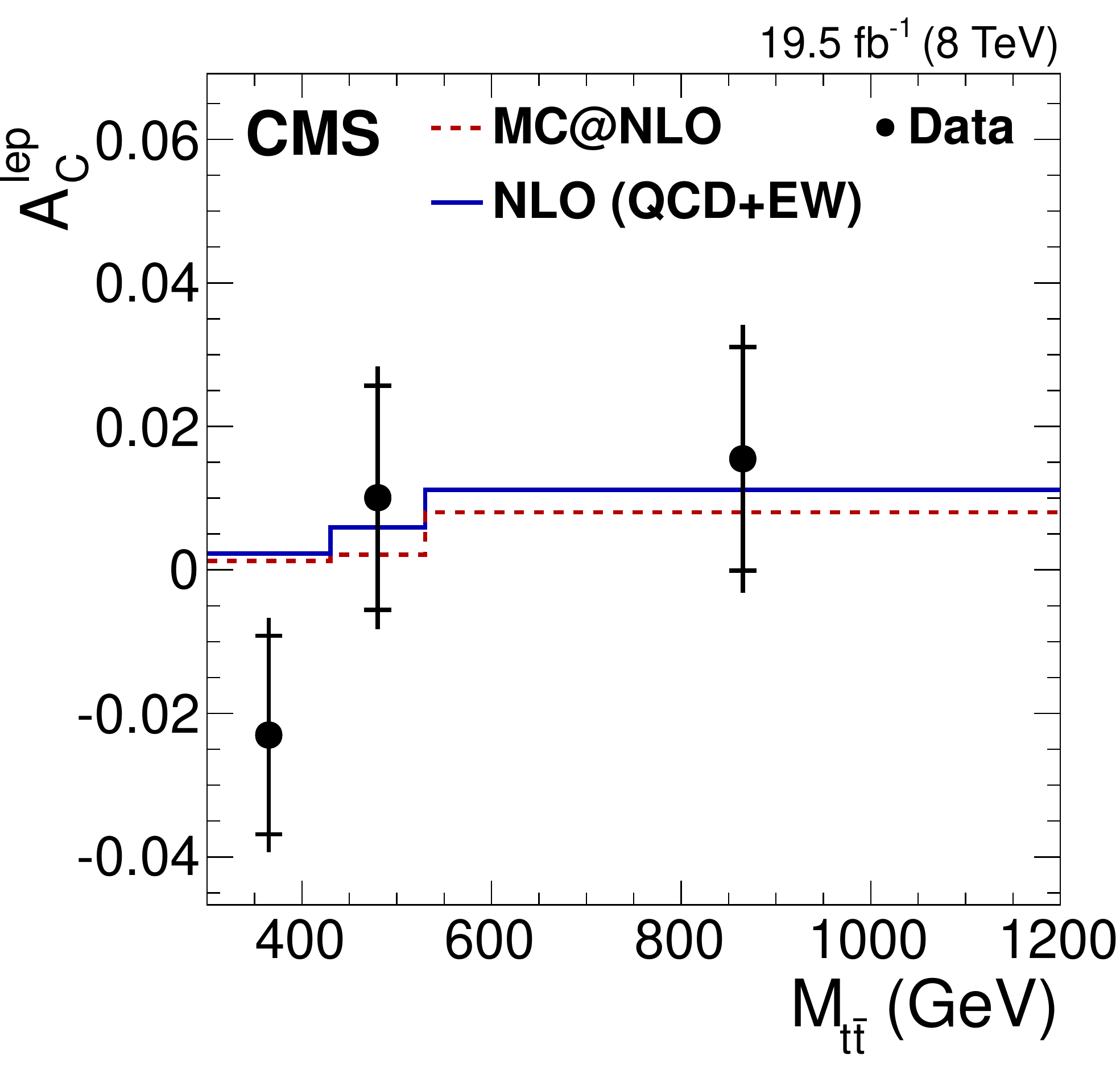}}
     \subfloat[]{\includegraphics[width=0.4\textwidth , height=0.31\textwidth]{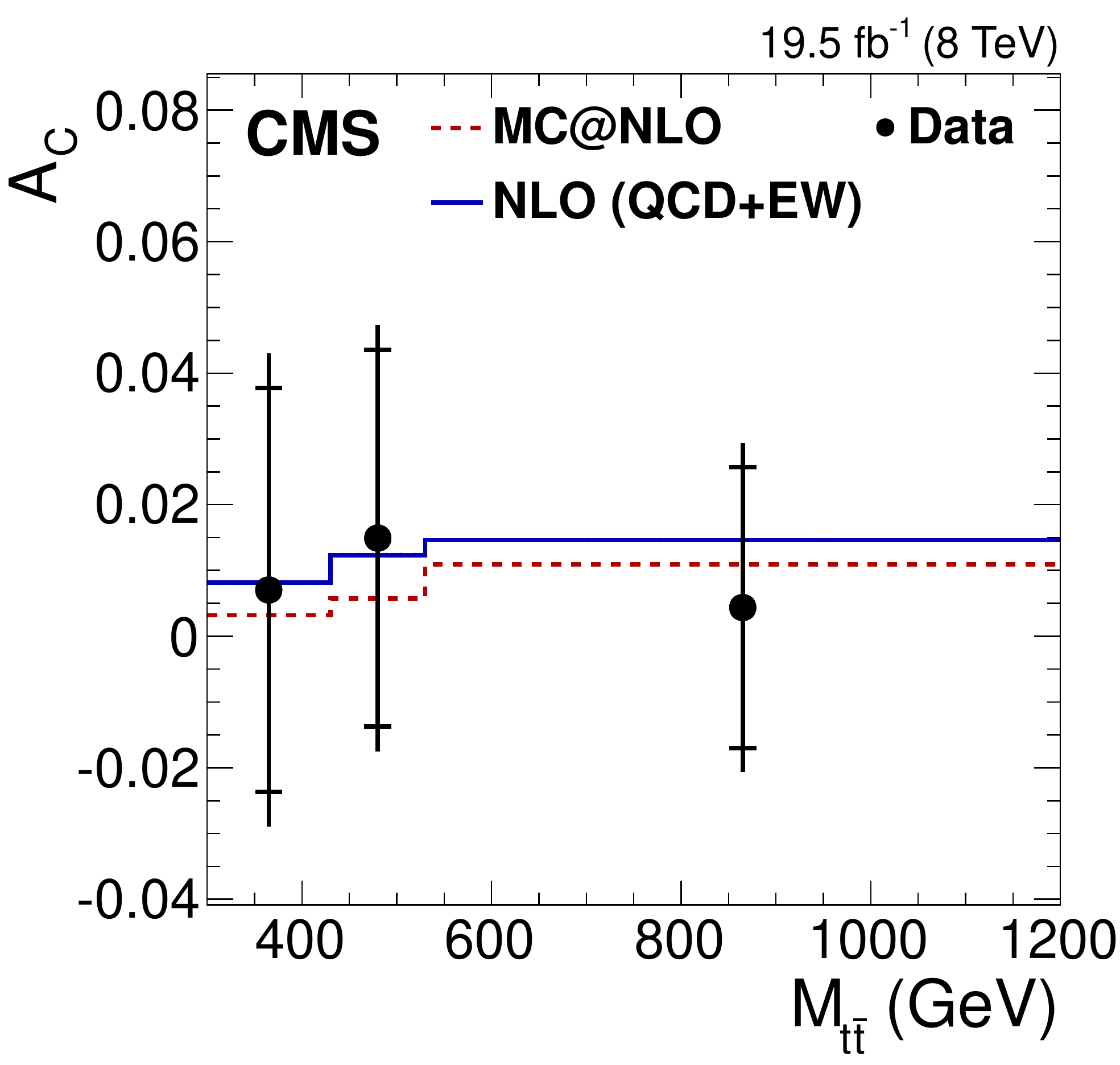}}\\
      \subfloat[]{\includegraphics[width=0.4\textwidth , height=0.3\textwidth]{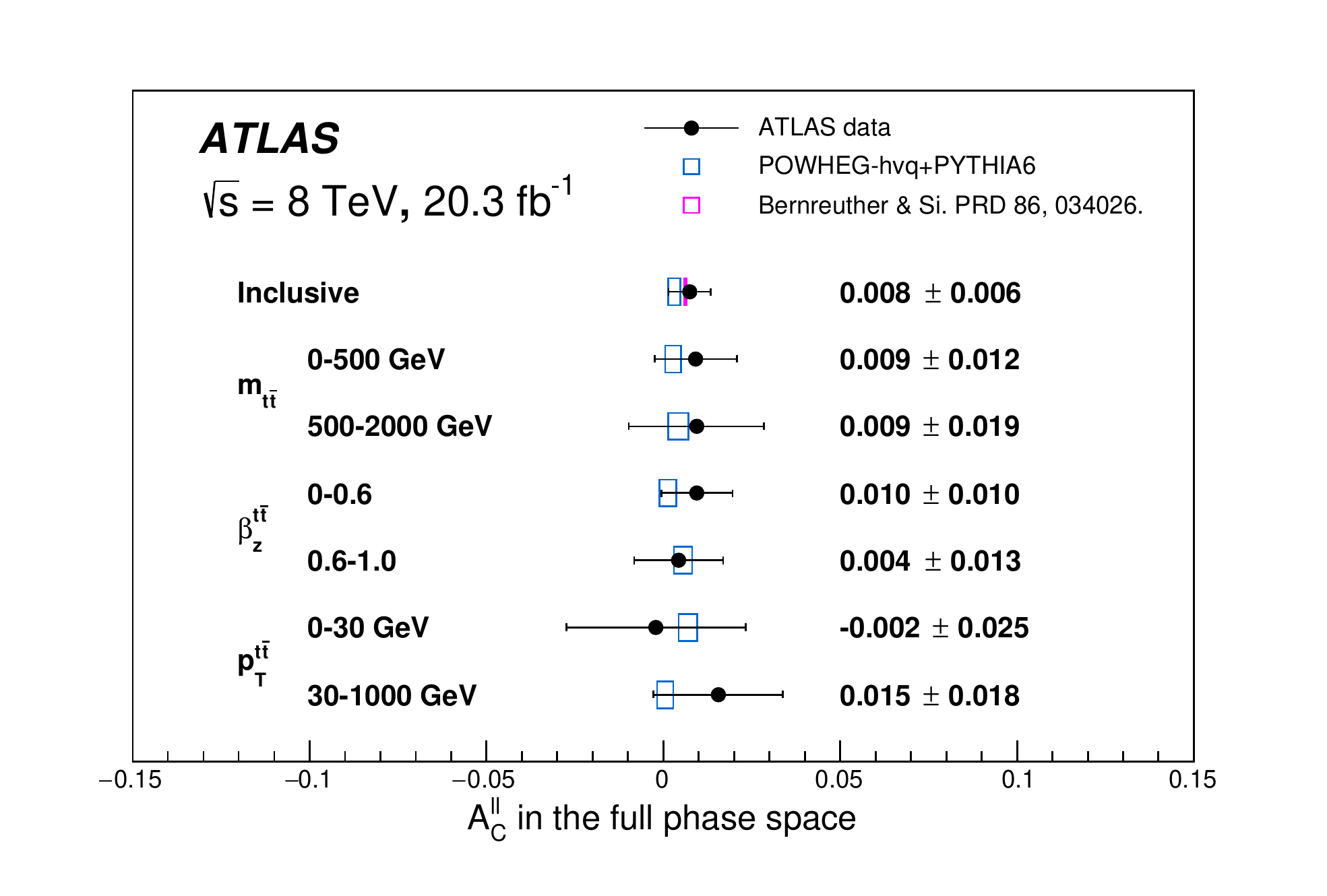}}
     \subfloat[]{\includegraphics[width=0.4\textwidth , height=0.3\textwidth]{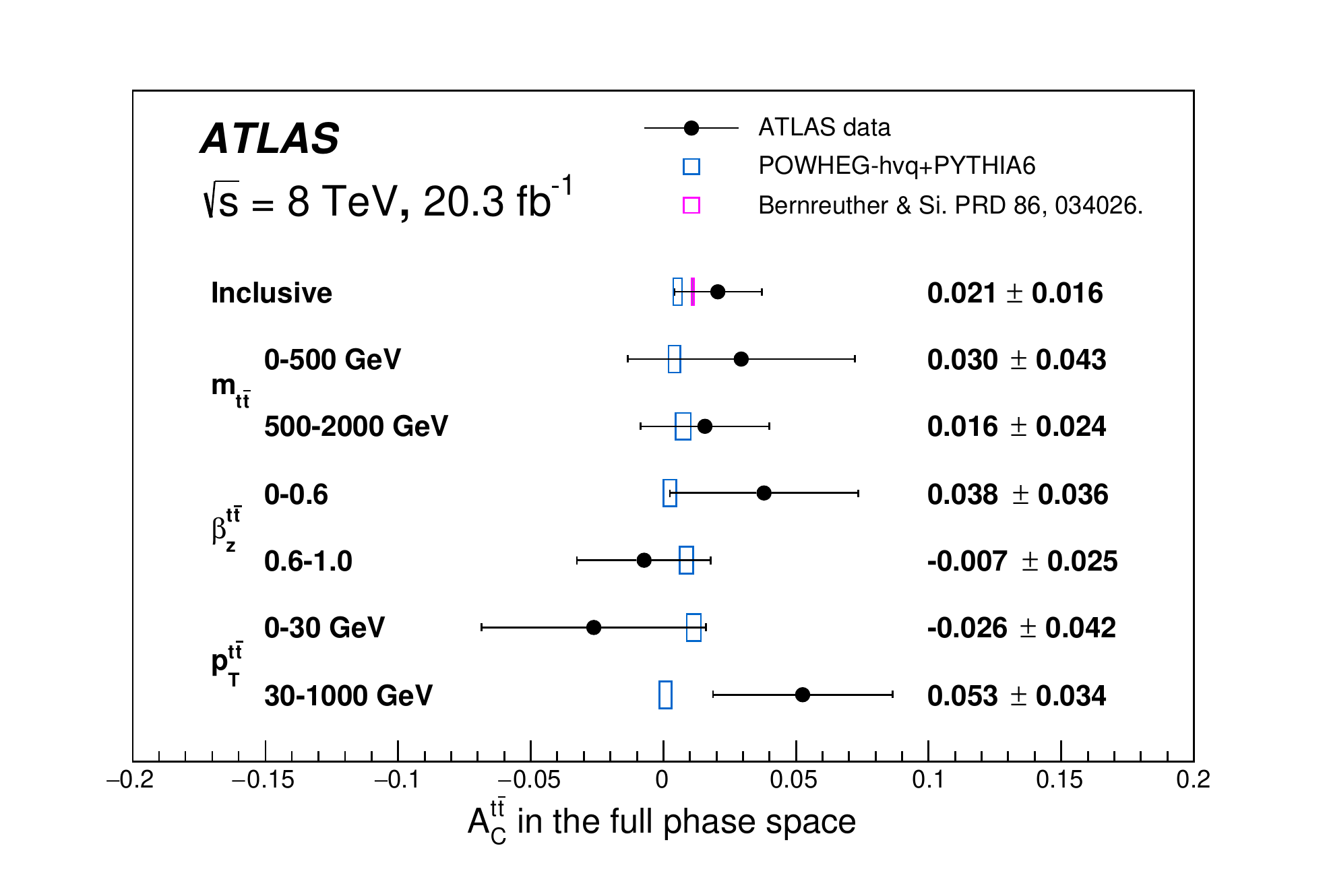}}
     \caption{(a,b): Dependence of the $t\bar{t}$ and leptonic charge asymmetries on $m_{t\bar{t}}$(CMS)~\cite{Khachatryan:2016ysn}. (c,d): Summary of the measurements  for the $t\bar{t}$ and leptonic asymmetries (ATLAS) ~\cite{Aad:2016ove}.\label{fig::dilep}}
\end{center}
\end{figure}

The CMS performs the same unfolding approach as smileptonic to measure the charge asymmetry in dilepton events with 19.7 fb$^{-1}$ accumulated data~\cite{Khachatryan:2016ysn}. The main benefit of lepton based asymmetry measurement is the absence of distorting effects in the $t\bar{t}$ reconstruction, yielding a smaller uncertainty after unfolding procedure. The lepton asymmetry is measured using $\Delta|\eta|$ of the two leptons. Figure~\ref{fig::dilep}(upper row) shows the charge asymmetry (right) and the lepton asymmetry (left) as a function of the invariant mass of the $t\bar{t}$ system. The measured inclusive asymmetries are A$_C$ = 0.011 $\pm$ 0.011(stat.) $\pm$ 0.007(syst.) and A$_{C}^{lep}$ = 0.003 $\pm$ 0.006(stat.) $\pm$ 0.003(syst.).


The ATLAS collaboration reports inclusive, and differential asymmetry as a function of the invariant mass, transverse momentum and longitudinal boost of the $t\bar{t}$ system(Figure~\ref{fig::dilep} lower row) ~\cite{Aad:2016ove}. The inclusive asymmetry are measured to be A$_C$ = 0.008 $\pm$ 0.006 for the lepton asymmetry and A$_C$ = 0.021 $\pm$ 0.016 for the $t\bar{t}$ asymmetry.
\section{The $t\bar{t}$ charge asymmetry in boosted events}
Apart from previous measurements, the ATLAS Collaboration analyzes lepton+jets events with a boosted topology, where the hadronic top-quark decay is reconstructed as a single large-radius jet~\cite{Aad:2015lgx}. The charge asymmetry is measured in a fiducial region with  $m_{t\bar{t}} > $  0.75 TeV and an absolute rapidity difference within $-2 < |y_{t}| - |y_{\bar{t}}| < 2$.
\begin{figure}[htb]
\begin{center}
     \subfloat[]{\includegraphics[width=0.45\textwidth , height=0.4\textwidth]{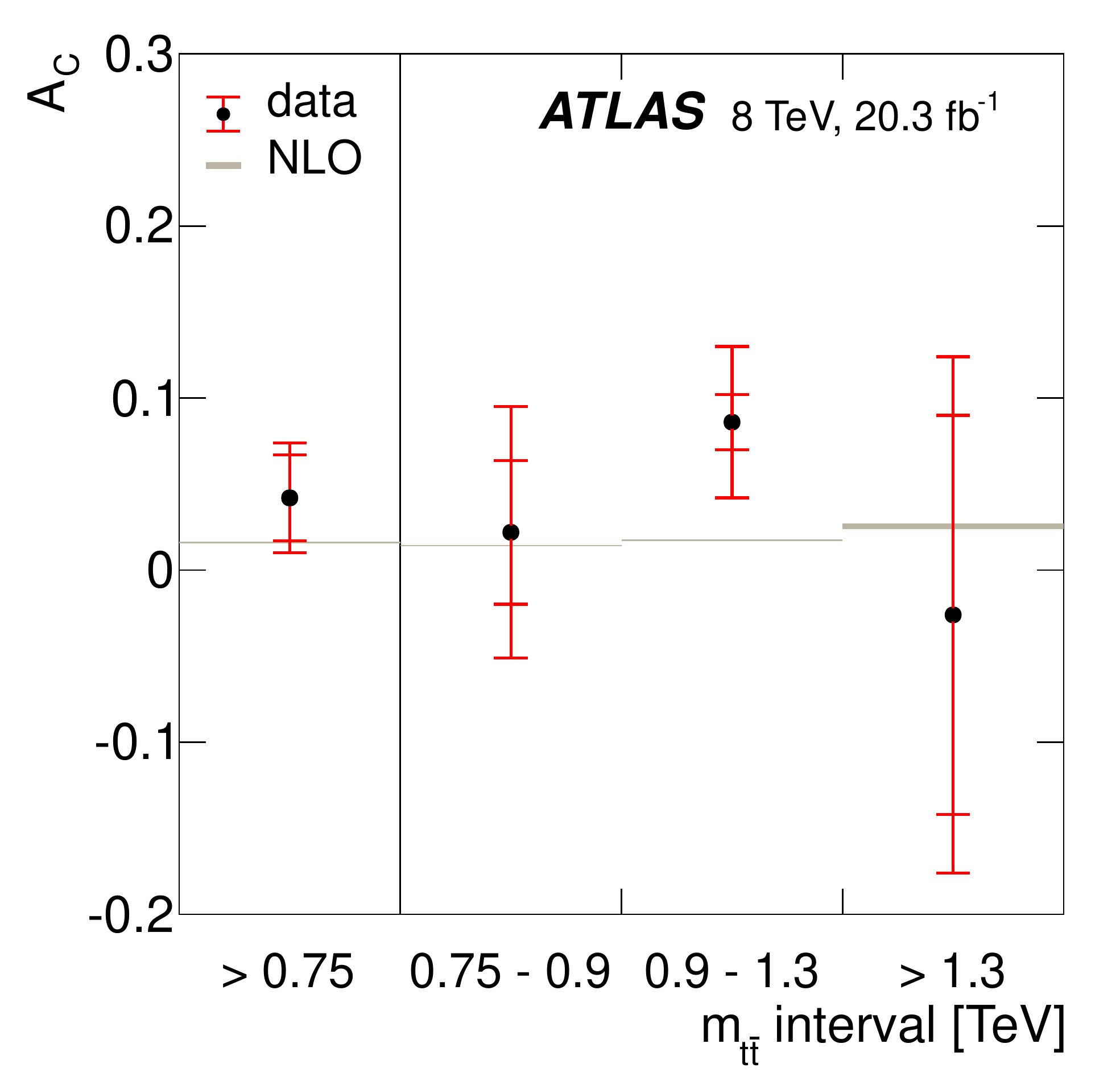}}
     \subfloat[]{\includegraphics[width=0.45\textwidth , height=0.4\textwidth]{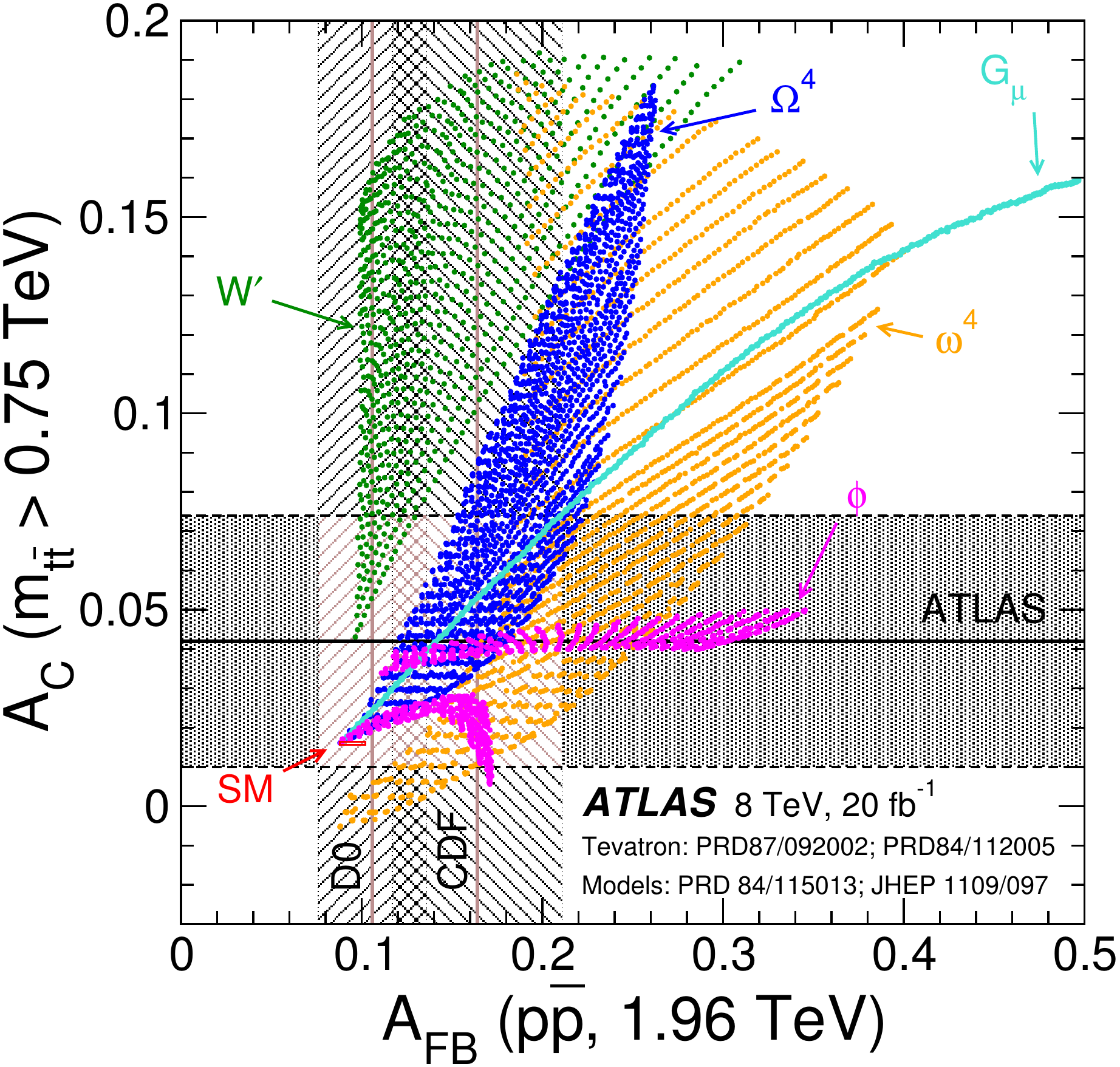}}
     \caption{(a) A summary of the charge asymmetry measurements in $m_{t\bar{t}}$  bins. (b) Predictions from a number of extensions of the SM for the forward-backward asymmetry, and high-mass charge asymmetry measurements at the LHC ~\cite{Aad:2015lgx}.\label{fig::boosted}}
\end{center}
\end{figure}
The measured inclusive asymmetry is  A$_C$ = 0.042 $\pm$ 0.032(stat $\oplus$ syst). Results of a differential measurement performed in three $m_{t\bar{t}}$  bins 
 are shown in Figure~\ref{fig::boosted}(left). Figure~\ref{fig::boosted} (right) compares the results of measurements of the asymmetry at Tevatron and LHC with their uncertainties. 
\section{The CP violation asymmetry at CMS}
\begin{figure}[htb]
\begin{center}
     \includegraphics[width=0.48\textwidth , height=0.45\textwidth]{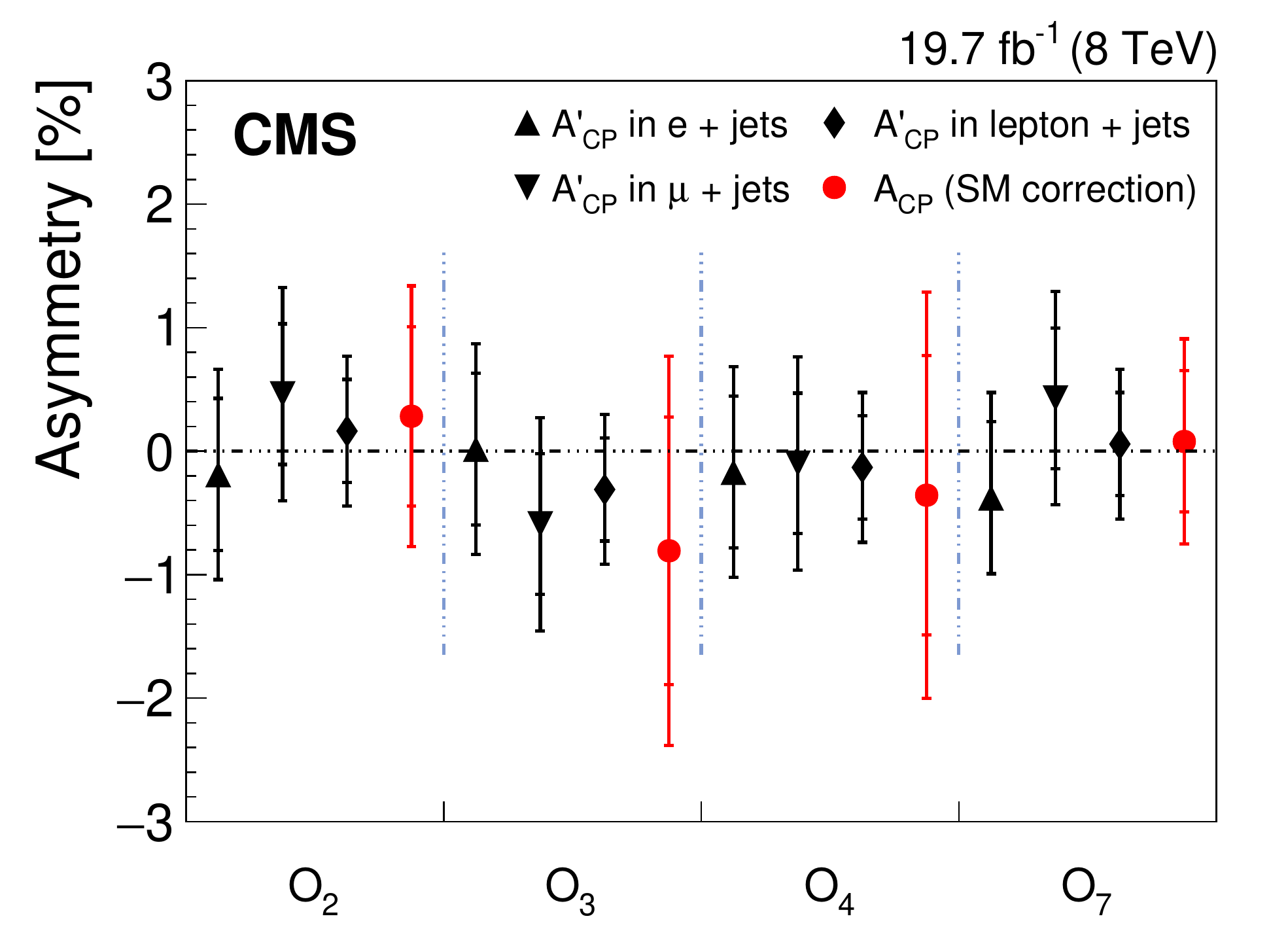}
     \caption{Summary of the uncorrected (corrected) CP asymmetries $A_{CP}(A_{CP}^{'})$ for the observables defined in~\cite{Khachatryan:2016ngh}.\label{fig::CMSCP}}
\end{center}
\end{figure}

The first measurement of CP violation asymmetries in top quark pair production and decay are performed in CMS with lepton+jets events~\cite{Khachatryan:2016ngh}. The analysis uses data from 8 TeV pp collisions corresponding to a total integrated luminosity of 19.7 fb$^{-1}$. Several new observables, as proposed in~\cite{Khachatryan:2016ngh}, are included to measure the CP violation asymmetries. These observables are odd under T transformation,  i.e. CP(O$_{i}$) =  -O$_{i}$. The A$_{CP}$ is found to be consistent with zero, within its uncertainty, in agreement with the standard model prediction as shown in Figure~\ref{fig::CMSCP}.
\section{Charge and CP asymmetries in b-hadron decays}
The ATLAS collaboration measures the same- and opposite-sign charge asymmetries in lepton+jets  $t\bar{t}$ events where a b-hadron decays semileptonically to a soft muon~\cite{Aaboud:2016bmk}. The charge asymmetries are formed based on the charge of the lepton from the top-quark decay and the charge of the soft muon from the semileptonic decay of a b-hadron. These asymmetries are measured in a fiducial region corresponding to the experimental acceptance. The data are categorized into same- and different-top-like SMT muons by a kinematic likelihood fitter (KLFitter)~\cite{Erdmann:2013rxa}. Given the charge asymmetries, four CP asymmetries (one mixing and three direct) are also measured.
\\
\section{Summary}
The latest results of $t\bar{t}$ asymmetry measurements performed by the CMS and ATLAS collaborations are presented.  All results are comparable with the predictions by the SM and no hint that point to new physics beyond the standard model is found.


\begin{thebibliography}{99}



\bibitem{Chatrchyan:2008aa} 
  S.~Chatrchyan {\it et al.} [CMS Collaboration],
  JINST {\bf 3}, S08004 (2008).
  doi:10.1088/1748-0221/3/08/S08004
    
\bibitem{Khachatryan:2015oga} 
  V. Khachatryan {\it et al.} [CMS Collaboration],
  Phys.\ Lett.\ B {\bf 757}, 154 (2016)
  doi:10.1016/j.physletb.2016.03.060
  [arXiv:1507.03119 [hep-ex]].

\bibitem{Khachatryan:2015mna} 
  V.~Khachatryan {\it et al.} [CMS Collaboration],
  Phys.\ Rev.\ D {\bf 93}, no. 3, 034014 (2016)
  doi:10.1103/PhysRevD.93.034014
  [arXiv:1508.03862 [hep-ex]].


\bibitem{Aad:2008zzm} 
  G.~Aad {\it et al.} [ATLAS Collaboration],
  JINST {\bf 3}, S08003 (2008).
  doi:10.1088/1748-0221/3/08/S08003


\bibitem{Aad:2015noh} 
  G.~Aad {\it et al.} [ATLAS Collaboration],
  Eur.\ Phys.\ J.\ C {\bf 76}, no. 2, 87 (2016)
  doi:10.1140/epjc/s10052-016-3910-6
  [arXiv:1509.02358 [hep-ex]].
 
\bibitem{Khachatryan:2016ysn} 
  V.~Khachatryan {\it et al.} [CMS Collaboration],
  Phys.\ Lett.\ B {\bf 760}, 365 (2016)
  doi:10.1016/j.physletb.2016.07.006
  [arXiv:1603.06221 [hep-ex]].
  
\bibitem{Aad:2016ove} 
  G.~Aad {\it et al.} [ATLAS Collaboration],
  Phys.\ Rev.\ D {\bf 94}, no. 3, 032006 (2016)
  doi:10.1103/PhysRevD.94.032006
  [arXiv:1604.05538 [hep-ex]].
  
  \bibitem{Aad:2015lgx} 
  G.~Aad {\it et al.} [ATLAS Collaboration],
  Phys.\ Lett.\ B {\bf 756}, 52 (2016)
  doi:10.1016/j.physletb.2016.02.055
  [arXiv:1512.06092 [hep-ex]].
  

\bibitem{Khachatryan:2016ngh} 
  V.~Khachatryan {\it et al.} [CMS Collaboration],
  [arXiv:1611.08931 [hep-ex]].
  
  
  \bibitem{Aaboud:2016bmk} 
  M.~Aaboud {\it et al.} [ATLAS Collaboration],
  JHEP {\bf 1702}, 071 (2017)
  doi:10.1007/JHEP02(2017)071
  [arXiv:1610.07869 [hep-ex]].
  
\bibitem{Erdmann:2013rxa} 
  J.~Erdmann, S.~Guindon, K.~Kroeninger, B.~Lemmer, O.~Nackenhorst, A.~Quadt and P.~Stolte,
  Nucl.\ Instrum.\ Meth.\ A {\bf 748}, 18 (2014)
  doi:10.1016/j.nima.2014.02.029
  [arXiv:1312.5595 [hep-ex]].
  
\end{thebibliography}
\end{document}